\documentclass[a4paper,11pt,times,astrosym]{aastex631}
\usepackage[english]{babel}
\usepackage{amsmath}
\usepackage{graphicx}
\usepackage{natbib}

\submitjournal{ApJ}

\shorttitle{The flow direction of interstellar neutral H from SOHO/SWAN} 

\newcommand{\kms}{~km~s$^{-1}$}

\newcommand{\lya}{Lyman-$\alpha$~{}}

\begin{document}

\title{The direction of the flow of interstellar neutral H based on photometric observations from SOHO/SWAN}

\correspondingauthor{M. Bzowski}
\email{bzowski@cbk.waw.pl} 

\author[0000-0003-3957-2359]{M. Bzowski}
\affil{Space Research Centre PAS (CBK PAN), Bartycka 18a, 00-716 Warsaw, Poland}

\author[0000-0002-5204-9645]{M.A. Kubiak}
\affil{Space Research Centre PAS (CBK PAN), Bartycka 18a, 00-716 Warsaw, Poland}

\author[0000-0003-3484-2970]{M. Strumik}
\affil{Space Research Centre PAS (CBK PAN), Bartycka 18a, 00-716 Warsaw, Poland}

\author[0000-0002-6569-3800]{I. Kowalska-Leszczynska}
\affil{Space Research Centre PAS (CBK PAN), Bartycka 18a, 00-716 Warsaw, Poland}

\author[0000-0001-8252-4104]{C. Porowski}
\affil{Space Research Centre PAS (CBK PAN), Bartycka 18a, 00-716 Warsaw, Poland}

\author[0000-0001-5376-2242]{E. Qu{\'e}merais}
\affil{LATMOS/IPSL, CNRS, UVSQ Universit{\'e} Paris-Saclay, Sorbonne Universit{\'e}s, Guyancourt, France}

\begin{abstract}
Interstellar neutral hydrogen flows into the heliosphere as a mixture of the primary and secondary populations from two somewhat different directions due to splitting occurring in the magnetized outer heliosheath. 
The direction of inflow of interstellar neutral H observed in the inner heliosphere, confronted with that of the unperturbed flow of interstellar neutral helium, is important for understanding the geometry of the distortion of the heliosphere from axial symmetry. 
It is also needed for facilitating remote-sensing studies of the solar wind structure based on observations of the helioglow, such as those presently performed by SOHO/SWAN, and in a near future by IMAP/GLOWS. 
In the past, the only means to measure the flow direction of interstellar hydrogen were  spectroscopic observations of the helioglow. 
Here, we propose a new method to determine this parameter based on a long series of photometric observations of the helioglow. 
The method is based on purely geometric considerations and does not depend on any model and absolute calibration of the measurements. We apply this method to sky maps of the helioglow available from the SOHO/SWAN experiment and derive the mean flow longitude of interstellar hydrogen. 
We obtain $253.1\degr \pm 2.8\degr$, which is in perfect agreement with the previously obtained results based on spectroscopic observations.
\end{abstract}

\section{Introduction}
\label{sec:intro}
\noindent
Interaction between the neutral and the charged, magnetized components of the local interstellar medium in front of the heliopause is one of the key processes shaping the heliosphere. 
Interstellar neutral hydrogen (ISN H) penetrates the heliopause and flows into the solar vicinity. 
A substantial portion of the ISN H population is ionized in front of the heliopause by charge exchange reactions with the ionized part of the interstellar matter that flows past the heliopause. 
As a result, a secondary population of ISN H is created, which is in a different physical state to that of the primary population and also penetrates inside the heliopause.  
Consequently, a superposition of the two ISN H populations inside the heliopause is expected to have distinctly different parameters to those of the unperturbed interstellar matter \citep[see, e.g.,][]{baranov_etal:91, baranov_malama:93}.

The heliosphere is distorted from axial symmetry by the action of interstellar magnetic field, as predicted by models \citep[e.g.,][]{ratkiewicz_etal:98a, ratkiewicz_etal:02a}, and verified by observations of the helioglow \citep{lallement_bertaux:90a,lallement_etal:93a, lallement_etal:05a} and energetic neutral atoms \citep{zirnstein_etal:16c}. 
As a result, the primary and secondary populations of ISN H flow into the heliosphere from directions different by several degrees from each other. 
On the average, the inflow direction for ISN H differs from that of the Sun's motion through the local interstellar medium, which is known from direct-sampling observations of ISN He \citep{witte:04, bzowski_etal:14a, bzowski_etal:15a, swaczyna_etal:22b}.
Determining the inflow direction of ISN H is thus important, because when compared with that of ISN He, it allows to estimate the strength and direction of interstellar magnetic field \citep{lallement_etal:05a}. 

ISN H, and in particular the heliospheric resonant backscatter glow \citep{wallis:74a}, have been used for remote-sensing of the solar wind structure. 
The idea, put forward by \citet{joselyn_holzer:75}, is that the latitudinally-structured solar wind is responsible for a latitudinally-structured ionization of ISN H inside the heliosphere. 
As a result, the density distribution of ISN H bears imprints of the solar wind structure. 
ISN H is excited by the solar \lya emission and emits a fluorescent glow. 
The source function for this glow at a given locus inside the heliosphere is proportional to the local density of ISN H. 
Thus, when observed from 1 au, the helioglow bears imprints of the solar wind structure.
Consequently, the solar wind structure can be deduced from analysis of the helioglow distribution in the sky. 
This idea was successfully implemented by several authors \citep[e.g.,][]{bertaux_etal:96a,bertaux_etal:99, kyrola_etal:98, summanen_etal:93,summanen_etal:97,bzowski_etal:03a, lallement_etal:10a,katushkina_etal:13a,katushkina_etal:19a,koutroumpa_etal:19a}. 

An alternative method to infer the latitudinal structure of the solar wind is an analysis of interplanetary scintillation of remote compact radio sources \citep[e.g.,][]{kojima:79a, jackson_etal:97a, kojima_etal:98a, tokumaru_etal:12b, tokumaru_etal:21a, sokol_etal:13a, porowski_etal:22a, porowski_etal:23a}. 
The results obtained from these two methods are in general agreement, but some important discrepancies have been pointed out \citep{katushkina_etal:13a, katushkina_etal:19a}, which need to be understood. 

Retrieval of the solar wind structure from photometric helioglow observations is greatly facilitated when the flow parameters of ISN H can be considered as well known. Up to now, these parameters have been obtained mostly using spectroscopic observations available from SOHO/SWAN \citep{quemerais_etal:99,lallement_etal:05a, lallement_etal:10a}.
An alternative method would be welcome. 

Here, we develop a method of determining ecliptic longitude of the inflow direction of ISN H based on photometric observations of the helioglow at 1 au and apply it to observations from SOHO/SWAN, performed from the beginning of the mission in 1996 until 2021. 
This method is completely model- an calibration-independent and is based solely on the geometric analysis of a carefully selected subset of observations. 
It can be applied as well to available photometric observations from SOHO/SWAN as, in the future, to observations of the helioglow performed by the GLObal solar Wind Structure (GLOWS) experiment onboard the planned NASA mission Interstellar Boundary Explorer \citep[IMAP; ][]{mccomas_etal:18b}.

We start the discussion from derivation of the method based on an insight from a model of the helioglow (Section \ref{sec:methodDeriv}). 
This is only used to demonstrate and verify the idea of the measurement. 
Subsequently, we briefly discuss available SOHO/SWAN data and their preparation for use within our method (Section \ref{sec:data}).
With this, we present the processing of the data and derivation of the flow longitude (Section \ref{sec:longiDeriv}).
Discussion of the uncertainties and the sensitivity of the result to various parameters used in the derivation is presented in Section \ref{sec:discussion}.
The paper is concluded by a summary Section \ref{sec:summaConclu}.

\section{Derivation of the method}
\label{sec:methodDeriv}
\subsection{Baseline idea}
\label{sec:baselineIdea}
\noindent
The distribution of the helioglow in the sky varies with the vantage point, but it clearly depends on the spatial distribution of ISN H. 
This distribution is shaped by the solar gravity and radiation pressure in the \lya line on one hand, and on the other hand by the solar ionization factors, which include charge exchange between solar wind protons and ISN H atoms, and photoionization by solar EUV photons.
If the aforementioned solar emissions were invariable in time and spherically symmetric,   and the flow of ISN H was perfectly homogeneous far away in front of the Sun (''at infinity'') down to the heliopause, then the distribution of the density and flow velocity of ISN H inside the heliosphere would be axially symmetric relative to the Sun, with the symmetry axis being the flow direction. 
Then, the density and velocity of ISN H could be represented by the so-called hot model of interstellar gas, developed by \citet{meier:77a, fahr:78, fahr:79, lallement_etal:85b}; and \citet{wu_judge:79a}, and the intensity of the helioglow as observed from the Sun would also be axially symmetric.

Even in this idealized case, the sky distribution of the helioglow would differ depending on the distance from the Sun and the angular distance of the vantage point from the upwind direction. 
The axial symmetry of the helioglow would be obtained only for vantage points located precisely at the upwind or downwind axis.
In this case, the lightcurve obtained from scanning a Sun-centered circle in the sky would be flat, i.e., featureless. 

When the vantage point is shifted away from the symmetry axis, then the scanning circle centered at the Sun is no longer perpendicular to the flow axis and thus, the helioglow intensity sampled along this circle is no longer uniform. 
For a vantage point at a fixed distance from the Sun (e.g., 1 au), the magnitude of the maximum to minimum ratio for this observation geometry depends on the offset angle of the vantage point from the upwind direction: the greater the offset angle, the greater the ratio until the offset angle exceeds 90\degr.
Then, the ratio begins to decrease again to reach 1 when the  vantage point is located at the downwind axis.

Typically, helioglow observations are performed by spacecraft located close to the ecliptic plane. Since the flow direction of ISN H is inclined at a small angle to the ecliptic plane \citep{lallement_etal:05a}, an ecliptic-bound observer is never precisely at the upwind or downwind axis. 
Consequently, even for the idealized conditions defined above, a Sun-centered lightcurve is not expected to ever be flat. 
However, when an observer within the ecliptic plane is at the ecliptic longitude equal to that of the upwind or downwind direction, the maximum to minimum ratio obtained for a Sun-centered lightcurve will be the lowest.
Hence, when the longitude of the scanning circle for which the maximum/minimum ratio is the lowest, one can immediately adopt this longitude as the longitude of the flow direction of ISN H.

The baseline idea of the proposed method is the following: 
\begin{itemize} 
\item take daily observations of the helioglow performed by a detector traveling around the Sun in the ecliptic plane, like SOHO/SWAN, and in the future IMAP/GLOWS,
\item extract lightcurves for scanning circles centered at or near the Sun with a selected angular radius, 
\item calculate maximum to minimum ratios for these daily lightcurves, 
\item identify the longitudes for which the maximum to minimum ratios reach their semi-yearly minima, 
\item the longitudes of the scanning circles for which the minimum max/min intensity ratios are identified correspond to the longitudes of the downwind or upwind directions.
\end{itemize}

The proposed method allows to derive the flow longitude directly from the photometric maps of the helioglow. 
As we show further in the paper, it can be applied to non-idealized cases as well.
It does not rely on any modeling. 
It is also independent of the absolute calibration of the detector. 
The only requirement is that the calibration is stable during a calendar year and uniform spatially.
In the case of SOHO/SWAN, it requires that the two sensors of this instrument are reliably cross-calibrated, which has been shown to be true \citep{quemerais_etal:06b}. 
 
\subsection{Validation of the method by modeling}
\label{sec:modelValidation}
\noindent
To verify the idea outlined in Section \ref{sec:baselineIdea}, we performed a modeling study.
We simulated the lightcurves of the helioglow for vantage points close to the ecliptic plane distributed around the Sun approximately every two weeks. 
We calculated the maximum to minimum ratios for the simulated lightcurves and we searched for the longitues of the vantage points longitudes for which these ratios showed local minima. 
We verified that the longitudes of the max/min ratio minima agree with the assumed longitude of the inflow direction. 
We performed this study both for solar minimum and solar maximum conditions, and assuming that the ionization rate of ISN H and \textcolor[rgb]{0,0,0}{the solar illumination are}  either spherically symmetric (2D) or latitudinally structured (3D). 
Details of the simulations are presented in Section \ref{sec:simu}, and processing of the simulated lightcurves is discussed in Section \ref{sec:simuLCproc}.

With the simulations on hand, we followed the procedure of retrieval of the flow longitude from the virtual observations (Sections \ref{sec:modelValidation} and \ref{sec:ModelMacMinRatio}) and compared the results with the values adopted in the modeling setup.

With results of this idealized scenario on hand, we verified that departures from spherical symmetry of the simulated ionization rate (Section \ref{sec:latiStrucIoni}) and masking of portions of the lightcurves to prevent contamination from extraheliospheric sourced does not introduce bias in the retrieved flow longitude (Section \ref{sec:dealingWithMasks}).

\subsubsection{Simulations}
\label{sec:simu}
\noindent
The locations of the vantage points were selected from the set of the positions of SOHO/SWAN for which full-sky maps of the helioglow are available for the years 2009 (solar minimum) and 2014 (solar maximum).
We took a subset of these points distributed around the Sun in approximately 2-week intervals.
This interval corresponds to a half of the Carrington rotation period.
For these vantage points, we simulated the helioglow intensity along scan circles centered at 0\degr{} latitude and the longitudes equal to those of the Sun for the given day minus 4\degr.
The radius of the scanning circle was chosen 75\degr. 
This viewing geometry corresponds to that of the upcoming IMAP/GLOWS experiment. 

\begin{figure}[!ht]
\centering
\includegraphics[width=0.4\textwidth]{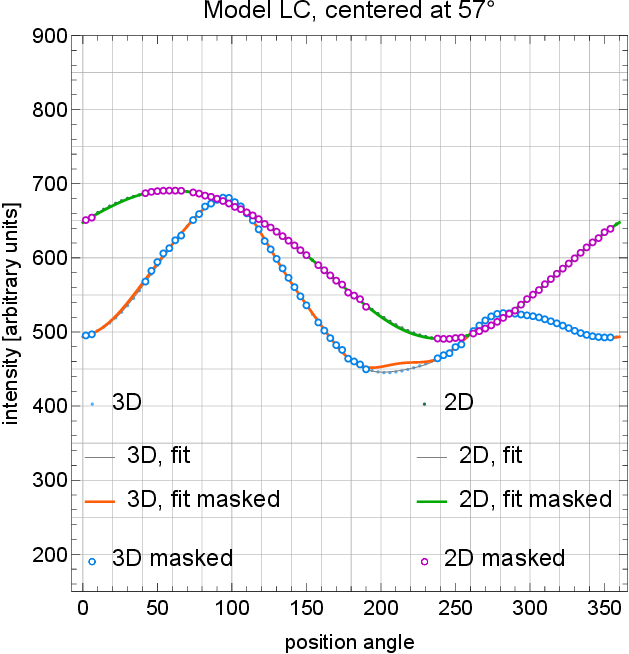}
\includegraphics[width=0.4\textwidth]{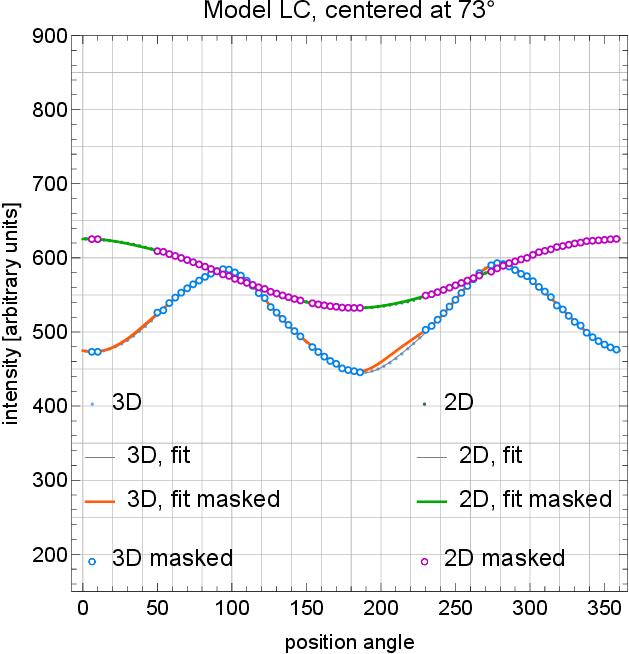}

\includegraphics[width=0.4\textwidth]{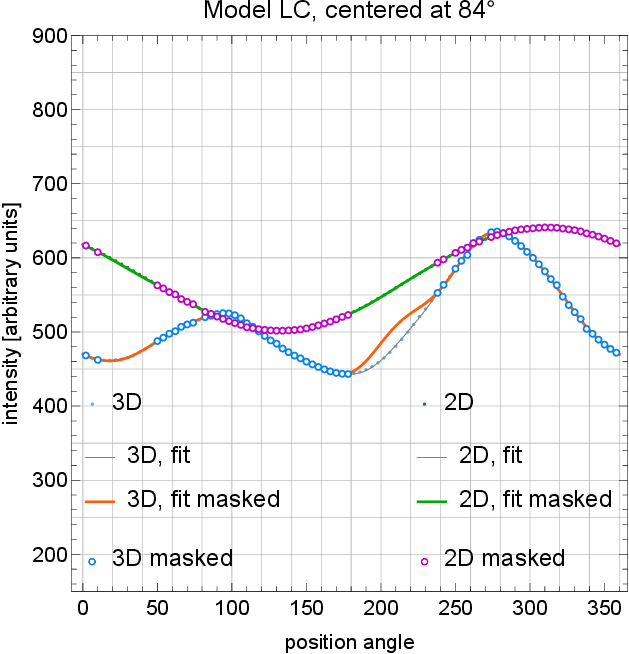}
\includegraphics[width=0.4\textwidth]{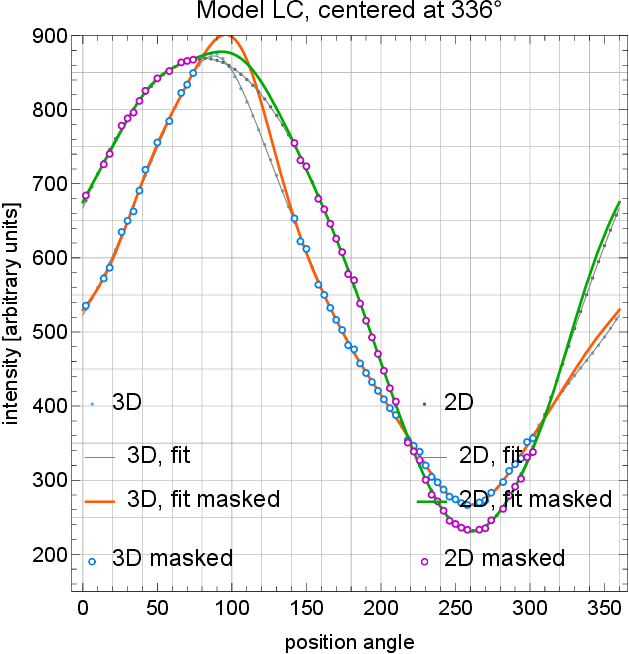}
\caption{Model lightcurves for selected vantage locations. The model was calculated for the solar wind and EUV output characteristic for 2014, i.e., for solar maximum conditions. 
A comparison of lightcurves for spherically symmetric (``2D'') and latitudinally-structured (``3D'') ionization rates is presented. 
Small dots represent lightcurves binned into 4\degr{} bins along the scanning circle. 
Open circles represent the same quantities left after application of SOHO/SWAN masks (see Figure \ref{fig:bafMaskPlot}). 
Solid lines represent fits of Equation (\ref{eq:sincosDef}) to full (thin lines) and masked (colored solid lines) light curves. The horizontal axis represents the position angle along the scanning circle counted off the northernmost point in the scanning circle. 
The numbers in the panel headers represent ecliptic longitudes of the centers of the scanning circles.
The lower-right panel represents a vantage point close to one of the crosswind positions, where the maximum to minimum ratio for a light curve is the greatest. 
The other three panels represent a passage of the observer through the upwind longitude. 
The vantage locations were selected to corresponds to those available in the SOHO/SWAN database (see Figure \ref{fig:swanLCSelPlots}).}
\label{fig:glcS2014plots}
\end{figure}

\begin{figure}[!ht]
\centering
\includegraphics[width=0.4\textwidth]{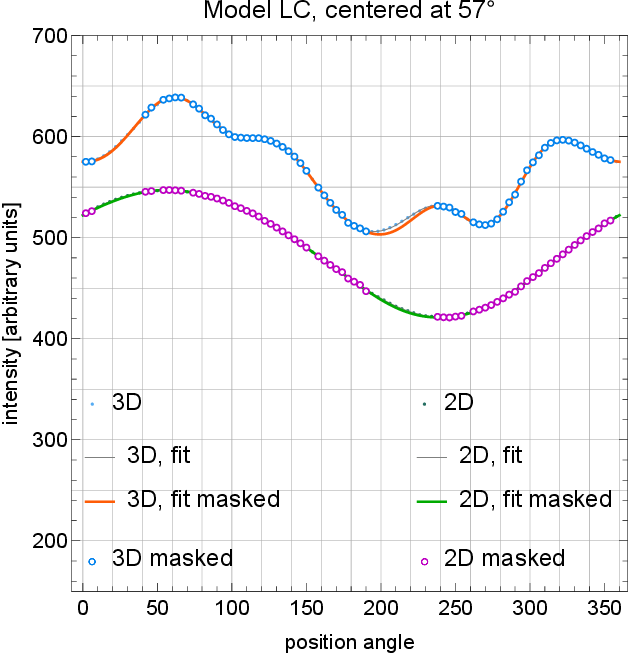}
\includegraphics[width=0.4\textwidth]{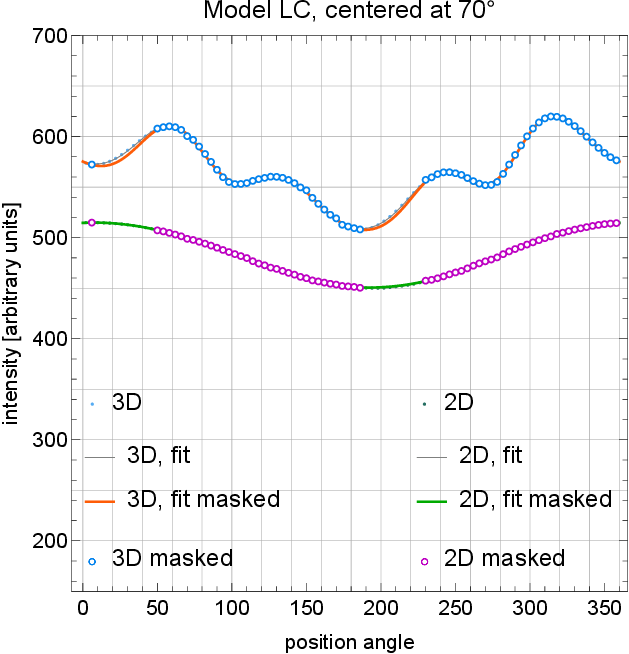}

\includegraphics[width=0.4\textwidth]{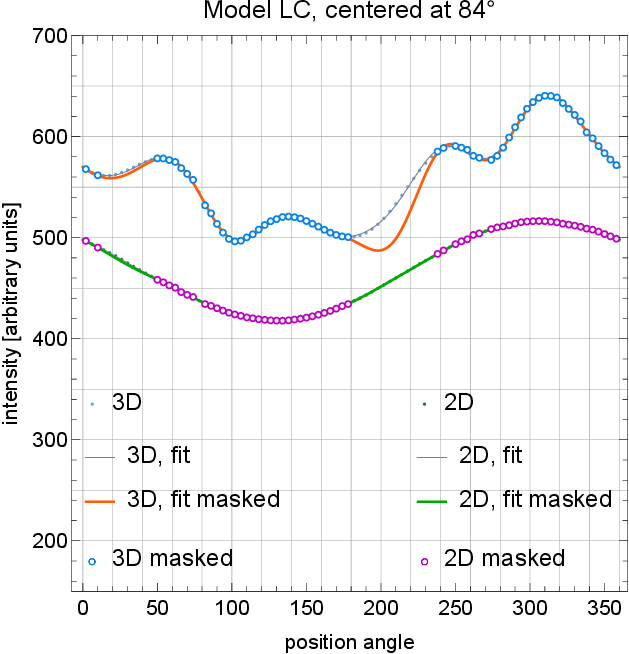}
\includegraphics[width=0.4\textwidth]{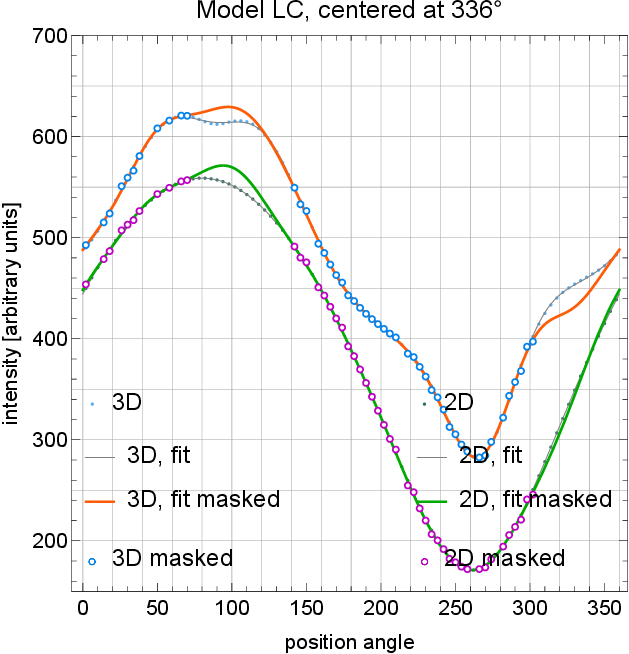}
\caption{Similar to Figure \ref{fig:glcS2014plots} but for 2009, characteristic for solar minimum conditions. }
\label{fig:glcS2009plots}
\end{figure}

The simulations were performed using the Warsaw Helioglow Model \citep[WawHelioGlow; ][]{kubiak_etal:21a, kubiak_etal:21b}.
For this part of the modeling, we assumed that the ionization rate of ISN H is spherically symmetric relative to the Sun but varies with time according to the predictions of the Warsaw Heliospheric Ionization Model \citep[WawHelioIon; ][]{porowski_etal:22a} for the ecliptic plane, and that radiation pressure is also spherically symmetric, but varies with time and radial velocity of individual H atoms \citep[WawHelioUV; ][]{IKL:20a}. 
These assumptions correspond to the cases marked as 2D in Figures \ref{fig:glcS2014plots} and \ref{fig:glcS2009plots}.

We assumed that ISN H inside the heliosphere is a mixture of two populations, the primary and the secondary. In the selection of the flow parameters, densities, and temperatures of these populations, we followed \citet{IKL:18b} (see their Table 1). 
The primary population at the heliopause is cooler, faster, and much less dense than the secondary. 
The inflow direction of the primary population conforms with that of the primary population of ISN He obtained by 
\citet{bzowski_etal:15a} based on IBEX observations of ISN He, and the secondary population flows from the inflow direction identical to that of the secondary population of ISN He, established by \citet{kubiak_etal:16a}. 

We simulated two series of 27 lightcurves for vantage points distributed approximately equally around the Sun. 
One of these simulation series was performed for the solar \lya flux evolving with time identically as in 2014, the other one for the conditions during 2009.
In Figures \ref{fig:glcS2014plots} and \ref{fig:glcS2009plots} (green lines, purple dots), we plot as an example the lightcurves for three vantage points chosen to be close to the upwind direction and for a point approximately 90\degr{} off the upwind direction.

The small blue and green dots in Figures \ref{fig:glcS2014plots} and \ref{fig:glcS2009plots} represent the example lightcurves, plotted as a function of the position angle along the scanning circle. 
The thin gray lines, marked ``2D, fit'' in the panels, represent functional approximations of the lightcurves, discussed in Section \ref{sec:simuLCproc}.

Inspection of the mentioned lightcurves supports the ideas presented at the beginning of Section \ref{sec:baselineIdea}. 
The greatest amplitude of the lightcurve is obtained for the crosswind vantage points, presented in the lower-right panel in Figures \ref{fig:glcS2014plots} and \ref{fig:glcS2009plots}. 
For the three other vantage points, the smallest amplitude is visible in the upper-right panel, which corresponds to vantage points the closest to the upwind longitude adopted in the simulation. 

\subsubsection{Processing of the lightcurves}
\label{sec:simuLCproc}
\noindent
Reliable calculation of the maximum to minimum ratio for a given lightcurve is best done by maximization and minimization of an approximation formula fitted to the simulations. 
We approximate the lightcurves using the following function:
\begin{equation}
F(\psi) = \frac{1}{\sqrt{\pi}}\left(\frac{1}{2}F_0 + \sum\limits_{k=1}^{n}\left(s_k \sin k \psi + c_k \cos k \psi \right) \right),
\label{eq:sincosDef}
\end{equation}
where $F_0, c_k, s_k$ are fit coefficients, obtained from fitting using a linear least-squares method, $\psi$ is the position angle along the scanning circle, and $n$ the order of the sin-cos series. 

This formula is a representation of the Fourier transform of a function. The order $n$ represents the cutoff and determines the scale of the smallest features that can be reproduced using this approximation. For example, for $n = 1$, the spatial scale is $\pi$, and for $n = 3$ the scale is $\pi/3$. Inspection of Figure \ref{fig:glcS2009plots} suggests that for representation of a lightcurve featuring four local maxima, an order $n \ge 4$ is needed. Generally, higher the order, the better the approximation; caveats will be discussed later in the paper.

For all lightcurves simulated within a given series (small gray dots in Figures \ref{fig:glcS2014plots} and \ref{fig:glcS2009plots}), we fit Equation \ref{eq:sincosDef} to the simulated lightcurve (see the thin gray lines in the example plots in Figures \ref{fig:glcS2014plots} and \ref{fig:glcS2009plots}), find numerically their maximum and minimum values, and calculate the maximum/minimum ratio. 
Along with the max/min ratio, we record the simulation time and ecliptic longitude of the scanning circle center. 
These latter longitudes are listed in the headers of the panels shown in Figures \ref{fig:glcS2014plots} and \ref{fig:glcS2009plots}. 

With all lightcurves from a simulation series processed, we have a series of the max/min ratios as a function of the scanning circle longitude, discussed in Section \ref{sec:ModelMacMinRatio}, which can be used to retrieve the upwind longitude of the flow direction of ISN H, as detailed in Section \ref{sec:upwindRetrieval}. 

\subsubsection{Effect of latitudinally-structured ionization rate}
\label{sec:latiStrucIoni}
\noindent
In reality, the solar wind, and consequently the ionization rate of ISN H, are latitudinally structured \citep[see][for most recent estimates]{porowski_etal:22a}. 
The helioglow distribution reflects this structure, featuring an ecliptic ``groove'' (a darkening) during solar minima, when the structuring is the most conspicuous. 
As a result, even for a vantage point at the ISN flow axis, a circumsolar lightcurve cannot be expected to be flat. An additional modulation of the lightcurve is added by the anisotropy in the illumination of the ISN gas by the latitudinally-anisotropic solar EUV emission.

We simulated helioglow lightcurves for identical boundary conditions and vantage points as those discussed in Section \ref{sec:simu} assuming the ionization rate model as obtained from the WawHelioIon model of the solar factors, that includes the charge exchange rate resulting from the solar wind structure obtained from IPS observations \citep{porowski_etal:22a} and the photoionization rate from \citet{sokol_etal:20a}. 
In this model, not only the solar wind, but also the photoionization rate and the solar illumination of the ISN H gas feature a small anisotropy \citep{strumik_etal:21b, kubiak_etal:21a}. The illumination anisotropy was adopted identical for the two simulation years, with the anisotropy factor equal to 0.85 \citep[see Equation 3.4 in ][]{bzowski_etal:13a}.

Selected model lightcurves and their fitted approximations are presented in Figures \ref{fig:glcS2014plots} and \ref{fig:glcS2009plots} (the cases marked as 3D). Clearly, they are much more structured than those simulated assuming a spherically symmetric ionization rate, but the conclusion that the maximum to minimum ratio becomes the lowest when the vantage point is at the longitude of the flow direction is supported. 
The processing of lightcurves simulated for the latitudinally-structured solar wind conditions (fitting the approximating function and retrieving the max/min ratio) is performed identically as discussed in Section \ref{sec:simuLCproc}. 
The locations of the maximum and minimum values in these lightcurves may be different to those in the case of spherically symmetric ionization model, but the general insight that the max/min ratio becomes smallest for vantage points at the upwind and downwind longitudes remains supported.

\subsubsection{Lightcurves with masks}
\label{sec:dealingWithMasks}
\noindent
In real world, some regions in the sky feature a strong extraheliospheric background, which basically precludes their use in the studies of the helioglow. 
\begin{figure}[!ht]
\centering
\includegraphics[width=0.5\textwidth]{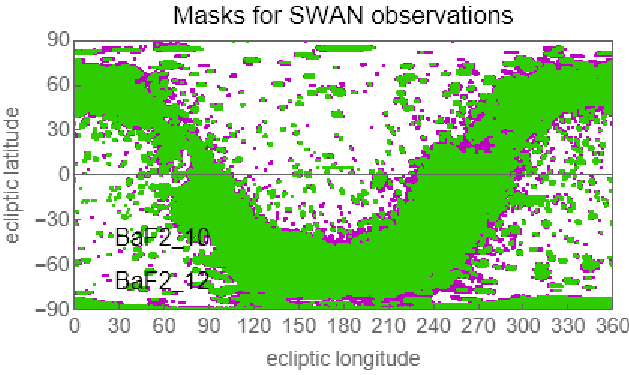}
\caption{Sky map in the ecliptic coordinates with two alternative masks for extraheliospheric background, defined by the SOHO/SWAN science team \citep{quemerais_etal:06b}.}
\label{fig:bafMaskPlot}
\end{figure}
\begin{figure}[!ht]
\centering
\includegraphics[width=0.5\textwidth]{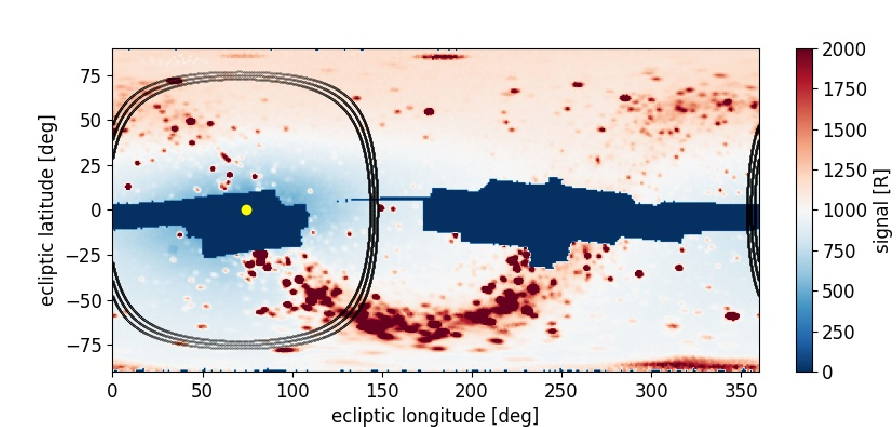}
\caption{Example full-sky map obtained from SOHO/SWAN on June 4, 2009, when the spacecraft was close to the upwind position. 
The axes are ecliptic longitude and latitude. 
The color scale marks the intensity of the helioglow. Dark-red features are extraheliospheric astrophysical objects, mostly stars. 
A band of these features corresponds to the Milky Way. 
The position of the Sun is given by the yellow dot, and the black ring corresponds to the adopted scanning circle, which has an angular radius of 75\degr{} and is centered at a point in the ecliptic plane with a longitude offset of 4\degr{} to the left of that of the Sun.
The lightcurve extracted from this map is presented in the upper-right panel in Figure \ref{fig:swanLCSelPlots}.
The dark-blue regions are portions of the sky obscured by the spacecraft body.
While the positions of extraheliospheric sources are fixed in the sky, the regions obscured by the spacecraft body differ between the vantage locations around the Sun.}
\label{fig:exampleSWANMap}
\end{figure}

This background is mostly due to EUV-bright stars and the Milky Way. 
Detailed contribution from these sources to the helioglow signal depends on the spectral sensitivity of individual instruments. 
Additionally, in some experiments, like SOHO/SWAN, a portion of the sky is obscured by the spacecraft body, and there is an avoidance region with a certain angular distance from the Sun that must be maintained to prevent instrument damage. 
This results in sky map masks that must be applied to the observed lightcurves. 
The science team of SOHO/SWAN published a mask for extraheliospheric sources, relevant for photometric observations from this experiment in \citet{quemerais_etal:06b}, which we plot in Figure \ref{fig:bafMaskPlot}. A mask for the spacecraft body for an example sky map is shown in Figure \ref{fig:exampleSWANMap}.

To verify that application of masks does not bias determination of the flow longitude, we repeated the procedure presented in Section \ref{sec:simuLCproc} for simulated lightcurves with the mask presented in Figure \ref{fig:bafMaskPlot} applied. 
Example lightcurves with the masks applied are shown in Figures \ref{fig:glcS2014plots} and \ref{fig:glcS2009plots} with open circles, and the fitted approximation functions are drawn with thick green lines for the spherically symmetric ionization case, and orange lines for the case of latitudinally structured ionization.  

In the case of lightcurves with masks, it sometimes happens that the region where the lightcurve has a maximum or a minimum, is masked. 
In these situations, we still obtain the maximum and minimum values from the fitted  approximation. 
However, the fitted model gives sometimes unrealistically high or low intensities but we verified that the region of the minima is almost always reproduced correctly. 

From analysis of unmasked lightcurves we found that for a yearly max/min ratio series, its maximum values is less than $\sim 4$. 
Thus, we restrict the allowable magnitude of the max/min ratio to 4. The lightcurves that after masking and fitting yield max/min ratios larger than 4 are eliminated from the sample.  

\subsection{Evolution of the maximum-to-minimum ratio during a calendar year}
\label{sec:ModelMacMinRatio}
\begin{figure}[!ht]
\centering
\includegraphics[width=0.45\textwidth]{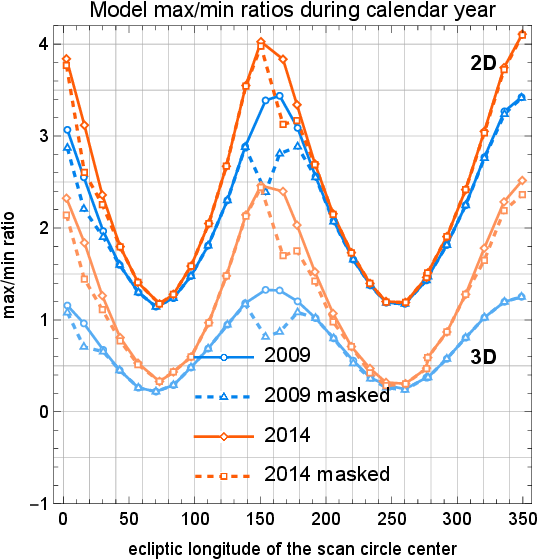}
\caption{Variation of the max/min ratios of the helioglow intensities obtained for lightcurves simulated for individual vantage points distributed around the Sun during two selected calendar years, corresponding to solar minimum (2009) and solar maximum conditions (2014). 
Solid lines represent model ratios obtained for full simulated lightcurves, and broken lines those for simulated lightcurves with SOHO/SWAN masks applied. 
The upper set of lines marked as 2D and drawn using more vivid colors, represents the ratios simulated adopting spherically symmetric ionization rate.
The lower set of lines, marked as 3D and drawn with pale colors, is offset by $-1$ relative to their native positions to improve the visibility.
They represent the ratios simulated adopting latitudinally-structured ionization rates. }
\label{fig:maxMinModelPlot}
\end{figure}

\noindent
The evolution of the model max/min ratios during a calendar year is presented in Figure \ref{fig:maxMinModelPlot}. The ratio features two local maxima and two minima. 
The positions of the minima correspond to the position of the center of the scanning circle the closest to the flow direction of ISN H adopted in the modeling. 
The magnitudes of the maxima of the ratios vary with the level of solar activity, as clearly demonstrated by the differences between the blue and orange lines in Figure \ref{fig:maxMinModelPlot}. 
Also, details of the adopted model of the ionization rate affect the amplitude of the max/min variation, as illustrated by comparison of the set of lines drawn with more intense and more pale colors.
The locations of the minima, however, are very stable and do not depend on the solar activity level and the latitudinal structure of the ionization rate.

Inspection of the ratios for the full lightcurves (solid lines) informs that the maximum of the max/min ratio rarely exceeds 4. 
The largest value of the maximum ratio is obtained for a spherically symmetric ionization rate and a high level of the solar activity.
During epochs of lower activity the maximum is lower, and introduction of a realistic latitudinal structure of the ionization rate reduces the height of this maximum even more.

Application of masks results in modified max/min series in their maximum regions. However, the magnitudes and locations of the minima are little affected, as demonstrated by comparisons of the ``masked'' ratios with their unmasked counterparts in Figure \ref{fig:maxMinModelPlot}.

\subsection{Retrieving the longitude of the inflow direction of ISN H}
\label{sec:upwindRetrieval}
\noindent
Retrieving the positions of the minima is done similarly as it was for the lightcurves themselves. 
The yearly series of the max/min ratios are fitted using the approximation function defined in Equation \ref{eq:sincosDef}, and the positions of the two minima are numerically searched for. 
From each calendar year, we obtain the downwind and upwind longitude. 
The downwind longitude is converted to the upwind longitude by addition of 180\degr{} to the result. 

The objective of the fitting is finding the phases of the minima, and not a precise reproduction of the shape of the max/min ratio. 
The max/min ratio features two yearly minima, and thus the minimum order in Equation \ref{eq:sincosDef} is $n = 2$. 
For the two simulated years and two cases of the ionization rate model for both of them, we investigated the results for the order $n$ in Equation \ref{eq:sincosDef} varying from 3 to 7. We performed this study for the cases of lightcurves with and without gaps. 
The results of the minimization show a slight variation with the order of the approximation formula defined in Equation \ref{eq:sincosDef}. 
We found that on the average, the retrieved upwind longitude was equal to 253\degr, varying between 251\degr{} and 254\degr. 
This is in excellent agreement with the assumed mean inflow direction equal to 253.21\degr{} (see Section \ref{sec:simu}).

Based on this study we concluded that the proposed method of determining the upwind longitude for ISN H is suitable for application to actual observations. This is discussed in the following section.   

\section{SWAN data and their processing}
\label{sec:data}
\subsection{Overview of the photometric data from SOHO/SWAN}
\label{sec:SWANdata}
\noindent
Solar Wind ANisotropies \citep[SWAN,][]{bertaux_etal:95} is an instrument on an ESA space mission Solar and Heliospheric Observatory \citep[SOHO; ][]{domingo_etal:95a} to observe the helioglow. 
For the purposes of this paper, we focus on photometric maps of the helioglow, collected every since the beginning of the mission in 1996. 

SOHO operates on an orbit around the solar-terrestrial L1 point. 
The vantage locations are close to those planned for the IMAP mission. 
The sky maps are collected using a twin periscope system, which scans two hemispheres of the sky with a narrow overlap between the fields of view of the southern and northern SWAN detectors to facilitate cross-calibration. 
During the first portion of the mission, maps were available every several days (with a break of several weeks in 1999 due to a mission disturbance), and after stopping spectroscopic observations with the use of a deuterium cell, they became available almost daily.
We used a total of 6172 sky maps. 
The data are publicly available at \url{http://swan.projet.latmos.ipsl.fr/data/L2//FSKMAPS/}. An example sky map is presented in Figure \ref{fig:exampleSWANMap}.

\begin{figure}[!ht]
\centering
\includegraphics[width=0.4\textwidth]{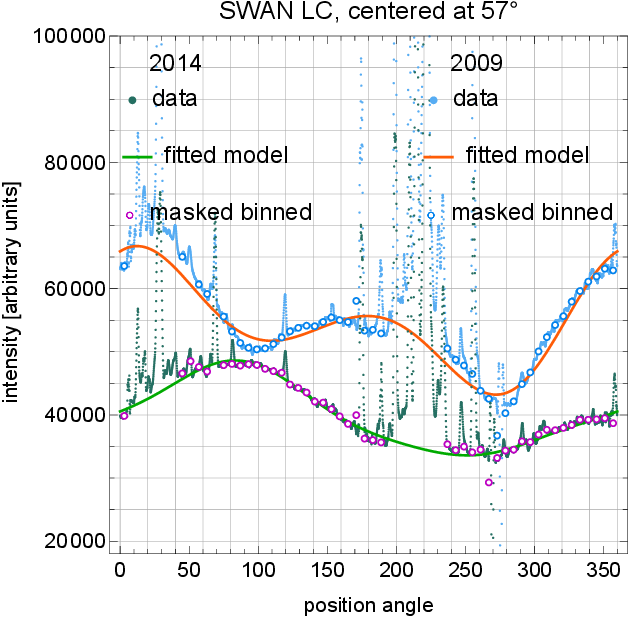}
\includegraphics[width=0.4\textwidth]{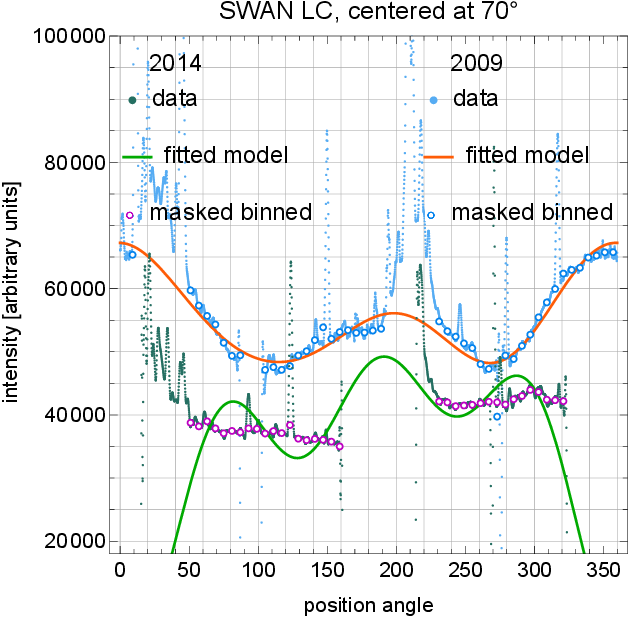}

\includegraphics[width=0.4\textwidth]{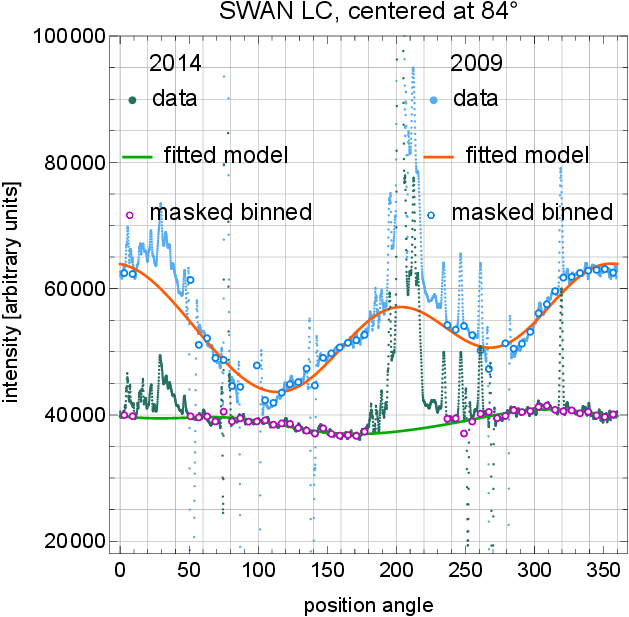}
\includegraphics[width=0.4\textwidth]{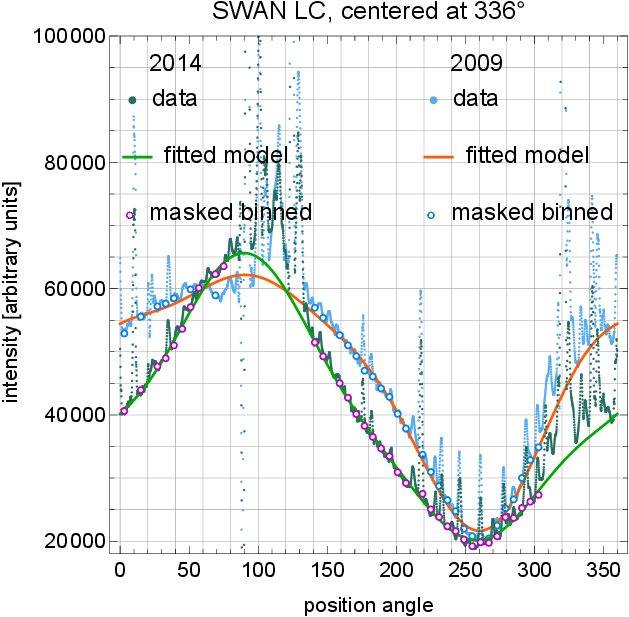}
\caption{Example lightcurves extracted from SWAN sky maps. Each of the panels presents data for selected vantage points observed in two calendar years, 2009 (the legends at the right side of the panels) and 2014 (the legends at the left side of the panels). 
The vantage points are identical to those used in Figures \ref{fig:glcS2014plots} and \ref{fig:glcS2009plots}.
Small dots, marked as ``data'', are lightcurves directly extracted from the maps. 
Open dots (``masked binned'') are the lightcurves with masks defined by \citet{quemerais_etal:06b} applied and binned into 6\degr{} bins in the position angle. 
Solid lines (``fitted model'') are the approximation functions defined in Equation (\ref{eq:sincosDef}) fitted to the masked and binned data. 
The horizontal axis is the position angle counted off the northernmost point of the adopted scanning circle. 
This circle intersects the ecliptic plane at 90\degr and 270\degr. 
Ecliptic longitudes of the centers of the scanning circles are listed at the top of individual panels. }
\label{fig:swanLCSelPlots}
\end{figure}

\subsection{Processing of the SWAN data before fitting the flow direction of ISN H}
\label{sec:dataProc}
\noindent
For this paper, from each of the available maps, we extracted a circular belt in the sky with an angular radius of 75\degr, centered at points in the ecliptic plane with longitudes lower by 4\degr{} than the longitudes of the Sun. 
In this way, we extract data for identical observation geometry to that planned for the IMAP/GLOWS experiment.
The published maps are sampled using bi-linear approximation to retrieve the measured intensities every 0.1\degr{} along the scanning circle. 

\begin{figure}[!ht]
\centering
\includegraphics[width=0.45\textwidth]{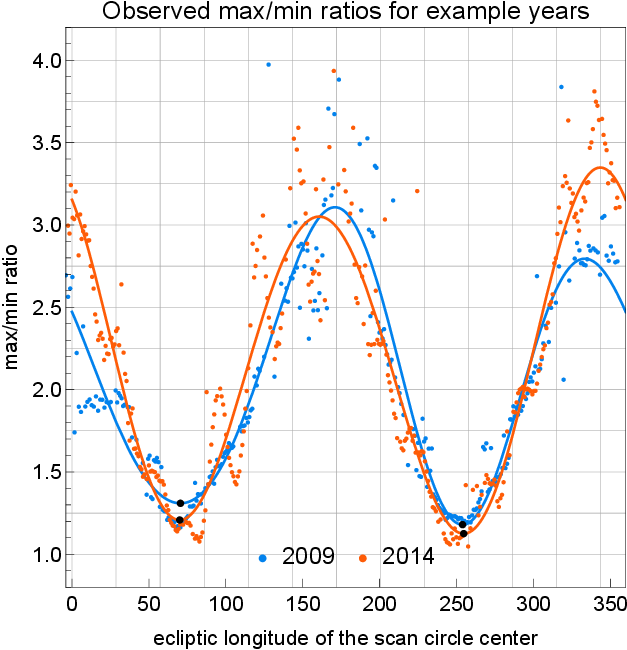}
\caption{Evolution of the observed max/min ratio during the two selected years, 2009 and 2014. Small dots represent the max/min ratios obtained from analysis of lightcurves retrieved from individual SWAN maps of the helioglow.
Solid lines mark the model fitted to the data.
Colors differentiate between the models and data for the calendar year 2009 (blue) and 2014 (orange).
The black dots mark the positions and magnitudes of the local minima obtained from analysis of the fitted models. 
The horizontal axis corresponds to ecliptic longitude of the can circle used to produces the light curve from an individual sky map, obtained from SOHO/SWAN observations.}
\label{fig:swanLCMinMaxSelPlot}
\end{figure}
In Figure \ref{fig:swanLCSelPlots}, we show as an example the observed lightcurves for the dates and vantage points close to those discussed in Section \ref{sec:simu}. The tiny dots present the lightcurves retrieved from the original SWAN maps. 
It is clearly seen that they have a substantial contribution from extraheliospheric astrophysical sources. 
The narrow spikes correspond to individual bright stars. 
The wider bands of enhanced emission, visible in the spin angle ranges of $\sim 180\degr - 240\degr$ and close to $\sim 30\degr$ in the upper row and the lower left panel, as well as close to $90\degr - 120\degr$ in the lower right panel, correspond to the Milky Way emission and bright stars seen against its background. 
These extraheliospheric emissions have fixed positions in the sky and, therefore, can be masked. 

The masks have been developed by the SWAN science team \citep{quemerais_etal:06b}.
We applied these masks to the lightcurves retrieved from all of the used maps and merged them with the masks corresponding to the body of the SOHO spacecraft (see the dark-blue regions in the example map shown in Figure \ref{fig:exampleSWANMap}).
Even after the masking, some contributions from extraheliospheric sources remains in the lightcurves. They make a fixed-pattern noise, which inevitably persists in the data.

The masked lightcurves were rebinned into 6\degr{} bins. 
Experiments we have made showed that adoption of this width of the binning is the best compromise between the need to maintain detail on one hand, and on the other hand, avoiding too much detail in the lightcurves due to the remaining unmasked extraheliospheric sources. 

The binning does not address the issues related to non-uniform solar emissions in the \lya{} line. 
These latter non-uniformities are responsible for the so-called search-light effect \citep{bertaux_etal:00}. 
A bright region at the solar surface illuminates a portion of the sky. 
The illuminated region moves eastward as the Sun rotates. 
A lightcurve such as that we deal with in this paper has a portion with an enhanced emission. 
The magnitude and spatial range of this effect in the context of lightcurves retrieved from SWAN maps were discussed by \citet{bzowski_etal:03a}. 
The amplitude varies between $\sim 1$\% and $\sim 5$\%, with the larger value characteristic for the solar activity maximum conditions.
The searchlight effect can potentially affect the derived max/min ratios for individual lightcurves and bias the flow longitude derived from a yearly set of observations.
A mitigation of this unwanted effect is performed statistically, by averaging the results obtained from multiple years of observations.

The masked and re-binned lightcurves are shown as purple and blue circles in Figure \ref{fig:swanLCSelPlots}. Clearly, they feature gaps.
Some of these gaps occupy a substantial portion of a lightcurve and occur in the regions where a maximum or minimum of the lightcurve can be expected.
With this, it is not possible to extract the magnitude of the minima and maxima of the lightcurves directly from the masked data product.
To address this issue, we approximated the lightcurves retrieved from the maps using the model defined in Equation (\ref{eq:sincosDef}) with $n = 3$. 
This prevents reproducing stars and asterisms, as well as the ``searchlights'' and permits to cover most of the masked regions, i.e., gaps in the lightcurves.
However, it does not allow to cover the largest ones. 
Adoption of a relatively large scale in the model $n = 3$, which corresponds to a scale of 60\degr{} in the position angle, results in smoothing over the inevitable jitter due to measurement background, individual stars, etc., visible in the example lightcurves in Figure \ref{fig:swanLCSelPlots}. 
Experiments showed that using a model with this relatively low order in comparison with a similar modeling performed on the simulated lightcurves discussed in Section \ref{sec:simuLCproc} reproduces the magnitudes and positions of the minima without reproducing smaller-scale features of the lightcurves related to the ``searchlights'' and contributions from unmasked extraheliospheric sources. 

Using the fitted lightcurve model, presented in the example Figure \ref{fig:swanLCSelPlots} as solid lines, we computed the magnitudes of the maximum and minimum values and calculated their ratios. 
The max/min ratios retrieved from all maps were split into series corresponding to individual calendar years.
Since on one hand, analysis of model lightcurves informed that ratios significantly exceeding 4 are not expected, and on the other hand gaps in the lightcurves fitted to the lightcurves retrieved from SWAN maps sometimes resulted in unrealistically high or low values of the approximating functions, we decided to reject the max/min ratios exceeding 4. 
One of the rejected cases is illustrated in the upper-right panel of Figure \ref{fig:swanLCSelPlots} (green line).

The series of max/min ratios corresponding to individual calendar years were fitted to Equation \ref{eq:sincosDef}. 
Each of these approximation functions features two local minima as a function of ecliptic longitude of the center of the scanning circle. 
These minima are separated by approximately 180\degr.
Their magnitudes correspond to ecliptic longitudes of the upwind or downwind directions.
Examples are shown in Figure \ref{fig:swanLCMinMaxSelPlot}.
For two selected years, 2009 and 2014, we show the max/min ratios of the lightcurves retrieved from SWAN maps for these years. The data feature a modulation as a function of ecliptic longitude of the center of the parent scanning circle. 
The fitted approximation functions are shown as solid lines, with black dots representing the magnitudes and longitudes of the local minima. 

A comparison of Figures \ref{fig:maxMinModelPlot} and \ref{fig:swanLCMinMaxSelPlot} indicates that the yearly variation of the max/min ratios obtained from the model is very similar to that obtained from the data. 

\section{Derivation of the inflow longitude}
\label{sec:longiDeriv}
\noindent
From the processing of the positions of local minima of the max/min ratios for individual years, we obtained two time series of the upwind longitude. 
Given the stability of the flow direction of ISN H reported by \citet{katushkina_etal:15a} and \citet{koutroumpa_etal:17a}, we did not expect any statistically significant changes in the inflow direction. 
Furthermore, we expected the difference between the positions of the two minima to be equal to 180\degr, at least on the average.
\begin{figure}[!ht]
\centering
\includegraphics[width=0.45\textwidth]{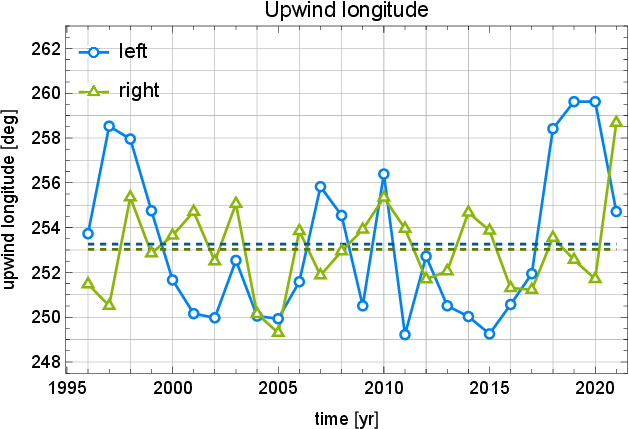}
\caption{Time series of the longitudes of the upwind direction, obtained from analysis of individual calendar years of SWAN maps. 
The blue line corresponds to the longitude obtained from the minima of the yearly max/min ratio at the left-hand side (scan circle longitude about 73\degr, marked as ``left''), the green line to that of the other minima. 
The two horizontal broken lines mark the mean values of the two respective time series. }
\label{fig:swanLCupLonPlot}
\end{figure}

Based on available analyses \citep{katushkina_etal:15a, koutroumpa_etal:17a}, a variation of the direction correlated with the phase of the solar activity cycle would be interpreted as an artefact of the method rather than a physical change in the local flow direction of the ISN H gas. 
Based on the insight on the time of flight of ISN atoms from the outer heliosheath to 1 au by \citet{bzowski_kubiak:20a}, we do not expect any detectable changes in the flow direction for fundamental reasons. 
The spread in the travel time of H atoms between the outer heliosheath and 1 au is larger than 30 years. 
Therefore, it does not seem feasible to detect a systematic secular change in the flow direction based on observations collected during a time span of $\sim 25$ years, even if, hypothetically, the Sun is traversing a region of interstellar gas with a gradient in the flow direction. 
Especially so, given the large random spread in the obtained flow directions due to variations of the observed lightcurves that originate deep inside the heliosphere (the searchlight effect and solar cycle-related variations).

With this in sight, given the relatively large scatter of the max/min ratios, we had to decide which approximation order to use in the approximation function given in Equation \ref{eq:sincosDef}. 
We performed a study with processing of the yearly max/min ratios in search for the minima while varying the order of the approximating function from 2 to 5. 

The most stable results were obtained for $n = 3$ (see Figure \ref{fig:swanLCupLonPlot}).
For this approximation order, we did not notice any significant correlation of the result with the phase of the solar cycle for the determination based on the right-hand minimum, and only moderate ones for the results obtained from the other minimum. 
The difference between the two mean upwind longitudes obtained from the two minima was the lowest (only 0.23\degr; see Figure \ref{fig:swanLCupLonPlot}). 

For $n = 2$, solar cycle-related effects were absent, but the difference between the two mean longitudes was 2.7\degr. 
For $n = 4$, some solar cycle periodicity in the results was noticed, and the difference between the results was $\sim 1.1\degr$. 
For $n = 5$, the difference increased to 5.7\degr, and a clear solar cycle periodicity was noticed.

Based on this study, we adopt the results obtained for $n = 3$. 
The averaged longitude of the upwind direction of ISN H was equal to $253.26\degr \pm 3.43\degr$ from the left-hand minimum and $253.03\degr \pm 2.00 \degr$ from the right-hand minimum. Collection of all the estimates from individual years regardless of the origin returns the mean upwind longitude equal to $253.1\degr \pm 2.8\degr$ (standard deviation of the full sample). 

\section{Discussion}
\label{sec:discussion}
\noindent
The results for ecliptic longitude of the upwind direction obtained using the method presented in this paper are robust. 
Even though the longitude series derived using a different order of the approximating function feature some unwanted effects, discussed in Section \ref{sec:longiDeriv}, the mean values are consistent with that adopted as the final result of the paper: the flow longitude is equal to 252.8\degr, 253.1\degr, 253.9\degr, 252.5\degr{} for $n = 2,\ldots, 5$, respectively. 
The scatter in these latter estimates is much less than the statistical uncertainty of the result obtained for $n = 3$.

The obtained inflow longitude agrees with expectations. 
Analyses of the flow direction of ISN H based on spectroscopic observations from SOHO/SWAN returned $252.3\degr \pm 0.7\degr$ \citep{quemerais_etal:99},  $252.5\degr \pm 0.7\degr$ \citep{lallement_etal:05a, lallement_etal:10a}, 251.8\degr{} \citep{katushkina_etal:15a}, $252.9\degr \pm 1.4\degr$ \citep{koutroumpa_etal:17a}; for a recent review of these studies, see Section 4.3.2 in \citet{baliukin_etal:22b}.
Our result, obtained from photometric, not spectroscopic observations, is in a very good agreement with these latter measurements. 

This result was obtained using a novel model-free method presented in this paper. The method relies solely on symmetry considerations. 
The derivation of this method was based on an insight from modeling, but modeling of the helioglow has not been used in the determination of the flow direction longitude.

In the analysis, we were unable to fully get rid of the influence of solar cycle modulation of the helioglow on the estimates of the flow longitude obtained for individual calendar years. 
However, since we used data from more than two full cycles of the solar activity, these effects were averaged out. 

The flow longitude derived from observations agrees very well with a model of the flow of ISN H based on direct-sampling measurements of the primary \citep{bzowski_etal:15a} and secondary populations \citep{kubiak_etal:16a} of ISN He, as well as an insight from global heliospheric modeling and Ulysses observations of pickup ions.
In this model, there are two populations of ISN H inside the heliosphere: the primary and the secondary. 
The inflow directions for these populations are adopted as identical to the flow directions of the primary and secondary ISN He, and the densities, speeds, and temperatures as those obtained by \citet{bzowski_etal:08a} based on analysis of the PUI production rate measurements on Ulysses, supported by modeling performed using the Moscow MC model \citep{izmodenov_etal:03a,izmodenov_etal:03b}. 
The average flow parameters can be obtained as respective weighted means of the respective parameters, with weights being the densities of the two populations, as we did in Section \ref{sec:simu}. 
The longitude of 253.21\degr{} thus obtained agrees excellently with the results of the model-free analysis presented in Section \ref{sec:longiDeriv}.

This two-population approach to modeling ISN H inside the heliosphere has been used in several studies \citep[e.g.,][]{IKL:18b, IKL:20a, IKL:22a, rahmanifard_etal:19a, kubiak_etal:21b}. 
It was also shown that for pickup-ion and helioglow studies, i.e., those where mostly the distribution of the zeroth- and first moments of the distribution function of ISN H are the most important, it is possible to simplify the modeling by the adoption of a one-population approach.
In this latter case, the parameters used by \citep{kubiak_etal:21b}: the (longitude, latitude) = $(252.5\degr, 8.9\degr)$, as measured by \citet{lallement_etal:10a}; the speed 21.26 \kms, obtained from weighted average of the Cartesian inflow velocity vectors of the primary and secondary populations; and the temperature $12\,860$ K, obtained from weighted average of the thermal speeds of the primary and secondary populations, are suitable to simplify modeling of the helioglow scanning circles observed by IMAP/GLOWS. 
Our model-independent measurement of the flow longitude lends support to this simplification.

\section{Summary and conclusions}
\label{sec:summaConclu}
\noindent
We have developed a new method of determination of the inflow direction longitude for ISN H. 
The method needs photometric observations of the helioglow performed from an approximately constant heliocentric distance along circles centered at a point located in the ecliptic plane within a few degrees from the Sun.
The observations should be performed during the entire calendar year, i.e., during a full revolution of the spacecraft around the Sun.
The method does not require any modeling.
It is based on an observation that the smallest maximum to minimum ratio for the daily lightcurves from Sun-centered small circles occurs when the detector is closest to the upwind or downwind direction.

Since in practice some portions of the daily lightcurves must be masked to eliminate contributions from the brightest extraheliospheric sources, like the Milky Way and brightest stars, it is convenient to approximate the daily lightcurves by a simple sine -- cosine series of a low order and to calculate the minimum to maximum ratios using the fitted approximate formula.

We have verified the applicability and robustness of the proposed method using simulations of the helioglow, performed using the Warsaw Heliospheric Glow Model (WawHelioGlow) for the solar maximum and minimum conditions, assuming various models of the ionization rate dependence on heliolatitude, and the inflow of two populations of ISN H (the primary and the secondary), entering the heliosphere from slightly different directions. 
In the simulations, we used realistic models of radiation pressure, illumination of ISN H by the solar \lya radiation, and of the ionization rate, based on recent models.

Application of the proposed method to simulated data reproduced ecliptic longitude of the inflow of ISN H within $\sim 0.2\degr$ from the effective inflow longitude assumed in the simulations.

With the proposed method verified, we have applied it to photometric observations of the helioglow available from SOHO/SWAN. 
We used available data from 1996--2021. 
We determined a yearly series of the inflow longitude estimates and based on these results, obtained for more than two cycles of the solar activity, we calculated the mean longitude of the inflow. 
We found it to be equal to $253.1\degr \pm 2.8\degr$, which is in excellent agreement with previous results derived based on spectroscopic observations of the helioglow.
We did not notice any temporal trends in the yearly time series of the inflow direction, which also agrees with expectations and previous studies.

A reliable estimate of the flow parameters of ISN H is important for the purposes of both the current SOHO/SWAN and the upcoming IMAP/GLOWS experiments, i.e., for remote-sensing studies of the solar wind. 
The latter one will perform helioglow observations with the observation geometry very close to that used in our paper. 
Its objective is to investigate the latitudinal structure of the solar wind and its evolution during the solar cycle. 
Accomplishing these objectives will require fitting the data with sophisticated simulations of the helioglow, with the inflow parameters as close to the reality as practical. 
This paper is an important inroad in this direction.

\noindent
Ackonwledgments\\
The work at CBK PAN was supported by Polish National Science Centre, grants 2019/35/B/ST9/01241 and 2018/31/D/ST9/02852, and Polish Ministry for Education and Science, contract MEiN/2021/2/DIR.

\bibliography{iplbib}
\bibliographystyle{aasjournal}  

\end{document}